\documentclass[11pt,onecolumn,romanappendices]{IEEEtran}

\usepackage{enumerate}
\usepackage[utf8]{inputenc} 
\usepackage[T1]{fontenc}
\usepackage{url}
\usepackage{ifthen}
\usepackage{tcolorbox}
\usepackage[noadjust]{cite}
\usepackage[cmex10]{amsmath} 
\usepackage{amssymb,amsfonts,amscd,amsbsy}
\usepackage{graphicx}
\usepackage{stmaryrd} 
\usepackage[mathscr]{eucal}
\usepackage{mathtools}
\usepackage{bbm}
\usepackage{xcolor}
\usepackage{xfrac}
\usepackage{algorithm}
\usepackage{algpseudocode}
\usepackage{enumerate}

\usepackage{hyperref}
\definecolor{cardinal}{rgb}{0.77, 0.12, 0.23}
\hypersetup{
    colorlinks=true,
    linkcolor=cardinal,
    citecolor=cardinal,
    urlcolor=cardinal
    }

\newtheorem{theorem}{Theorem}
\newtheorem{corollary}{Corollary}
\newtheorem{lemma}{Lemma}

\newtheorem{example}{Example}
\newtheorem{remark}{Remark}

\usepackage{etoolbox}
\AtEndEnvironment{example}{\null\hfill$\square$}%
\AtEndEnvironment{remark}{\null\hfill$\square$}%

\newcommand{\suppress}[1]{}

\newcommand{\Fb}{\mathbbmss{F}} 
\newcommand{\Cs}{\mathscr{C}}   

\newcommand{\Amin}{\mathcal{A}_{\min}}   
\newcommand{\supp}{\mathsf{supp}}  

\newcommand{\RM}{\mathrm{RM}}   
\newcommand{\tth}{\text{th}}    
\newcommand{\p}{\pmb}           

\newcommand{\dmin}{d_{\min}}    

\newcommand{\nosubscriptallone}{\mathbbm{1}}   
\newcommand{\allonelengthn}{\mathbbm{1}_n}   
\newcommand{\dimenof}{\mathsf{dim}}  
\newcommand{\rowspaceof}{\mathsf{rowsp}}  

\newcommand\code[2]{\mathscr{C}^{\otimes[#1,#2]}} 

\newcommand{\rank}{{\rm rank}}
\newcommand{\eval}{\p{\rm ev}} 
\newcommand{\Bs}{\mathscr{B}} 
\newcommand{\wtdist}{{\sf A}}

\title{Subcodes of Second-Order Reed-Muller Codes\\ via Recursive Subproducts}

\author{A P Vaideeswaran, Madireddi Sai Harish, Lakshmi Prasad Natarajan%
\thanks{\hrule}%
\thanks{The authors are with the Department of Electrical Engineering, Indian Institute of Technology Hyderabad, Sangareddy 502248, India (email: \{ee21btech11001, ee21btech11047, lakshminatarajan\}@iith.ac.in).}%
\thanks{This work was supported by the Qualcomm 6G University Research India Program.}
}

\begin{document}

\maketitle

\begin{abstract}
We use a simple construction called `recursive subproducts' (that is known to yield good codes of lengths $n^m$, $n \geq 3$) to identify a family of codes sandwiched between first-order and second-order Reed-Muller (RM) codes. 
These codes are subcodes of multidimensional product codes that use first-order RM codes as components.
We identify the minimum weight codewords of all the codes in this family, and numerically determine the weight distribution of some of them.
While these codes have the same minimum distance and a smaller rate than second-order RM codes, they have significantly fewer minimum weight codewords.
Further, these codes can be decoded via modifications to known RM decoders which yield codeword error rates within $0.25$~dB of second-order RM codes and better than CRC-aided Polar codes (in terms of $E_b/N_o$ for lengths $256, 512, 1024$), thereby offering rate adaptation options for RM codes in low-capacity scenarios.
\end{abstract}


\setcounter{tocdepth}{2} 
\tableofcontents
\clearpage


\section{Introduction}




%

\textbf{Subcodes of Second-Order Reed-Muller Codes.}
Emerging applications, such as the internet-of-things, have prompted interest in designing codes for channels with low capacity~\cite{FHJM_JSAIT_23,DuG_ISITA_2020,DuG_ISIT_2021}.
Low-rate Reed-Muller (RM) codes~\cite{Muller_IRE_54,Reed_IRE_54} are considered good candidates for low-capacity channels; in particular, RM codes with vanishing rates achieve arbitrarily high fractions of channel capacity in the binary erasure and the binary symmetric channels~\cite{ASW_IT_2015}. 
RM codes (with low rate and short length) are also used for encoding uplink control information in 5G New Radio.
However, for a given length \mbox{$N=2^m$}, RM codes provide only \mbox{$m+1$} distinct choices of code dimension $\sum_{i=0}^{r} \binom{m}{i}$, corresponding to order $r \in \{0,1,\dots,m\}$, which is an inconvenience at short and medium block lengths\footnote{We note that Polar codes~\cite{Arikan_IT_2009} are flexible in terms of code rate, but unlike RM codes, their construction is neither channel-independent nor explicit.}. 
Identifying good subcodes of RM codes can increase the options available for code rate, and can make RM codes attractive in terms of rate adaptation in practice.
Further, it will be advantageous if the decoding algorithm and the codeword error rate (CER) versus $E_b/N_o$ performance of these subcodes are similar to those of RM codes.
In this paper, we use a simple construction to identify one such family of subcodes of second-order RM codes.



Algebraic constructions of good subcodes of second-order RM codes exist, see Chapter~15 of~\cite{macwilliams1977theory}. 
For example, there exist subcodes with distance $\sfrac{3N}{8}$ that encode messages of length $\sfrac{1}{2} \log_2^2 N - \sfrac{1}{2} \log_2 N + 1$ (for odd values of $\log_2 N$) or $\sfrac{1}{2} \log_2^2 N - \sfrac{1}{2} \log_2 N + 2$ (for even values of $\log_2 N$), see Corollary~17 and Theorem~19 of~\cite[Chap.~15]{macwilliams1977theory}.
The latter codes (for even values of $\log_2 N$) are instances of the non-linear Delsarte-Goethals codes.
However, to the best of our knowledge, efficient soft-input decoders are not known for these codes.


Another way to construct subcodes of $\RM(2,m)$ of a specified dimension $K$ (albeit with distance only $\sfrac{N}{4}$), is to remove \mbox{$\dim(\RM(2,m))-K$} rows from the standard generator matrix of $\RM(2,m)$ to get the generator matrix of a subcode~\cite{TTK_IEICE_98,MHU_TCOM_2014,VABM_SCVT_2016,VBM_COMML_2018}.
In general, it is not known which choice of rows (to be removed) leads to a subcode with the smallest CER.
%
The works~\cite{CJL_COMML_19} and~\cite{JFMH_ICC_22} too consider RM subcodes obtained by removing certain rows from the generator matrix of RM codes. In particular,~\cite{CJL_COMML_19,JFMH_ICC_22} 
propose decoding algorithms for product codes whose component codes are RM codes;
note that the product $\RM(1,m_1) \otimes \RM(1,m_2)$ is a subcode of $\RM(2,m_1+m_2)$~\cite{SaA_IT_2005}, with distance $\sfrac{N}{4}$ and dimension $(m_1+1)(m_2+1) \leq \sfrac{1}{4} \log_2^2 N + \log_2 N + 1$, where $N=2^{m_1+m_2}$.




\textbf{Recursive Subproduct Codes.}
In this paper we use the construction of \emph{recursive subproduct codes} that is known to yield codes of lengths $n^m$, where $n \geq 3$ and $m \geq 2$, with CER close to that of RM codes~\cite{SNK_ISIT24}.
The construction of~\cite{SNK_ISIT24} yields subcodes of the $m$-fold product code~\cite{Elias_IRE_54} $\Cs^{\otimes m} = \Cs \otimes \cdots \otimes \Cs$ that share several properties of RM codes (Plotkin-like design, `projection' operation~\cite{YeA_IT_20}) and includes RM codes as a special case ($\Cs=\Fb_2^2$). 
%
%
The ingredients for constructing a recursive subproduct code are: a~\emph{base code} $\Cs$ that contains the all-ones codeword (and is of dimension at least $2$), and integers \mbox{$0 \leq r \leq m$}. The corresponding recursive subproduct code is denoted as $\code{r}{m}$ and $r$ is called the order. If the parameters of $\Cs$ are $[n,k,d]$, then the parameters of $\code{r}{m}$ are 
\begin{equation} \label{eq:basic_parameters_recursive_subproduct_codes}
\left[ n^m, \, \sum_{i=0}^{r}\binom{m}{i}(k-1)^i, \, d^r n^{m-r} \right].
\end{equation}
For any $m \geq 1$, $\code{0}{m} \subset \code{1}{m} \subset \cdots \subset \code{m}{m} = \Cs^{\otimes m}$.

\textbf{Contributions and Organization.}
In this paper we study recursive subproduct codes $\code{r}{m}$ with \mbox{$r=2$} and \mbox{$\Cs=\RM(1,m')$} for \mbox{$m,m' \geq 2$}. These codes are sandwiched between first- and second-order RM codes, i.e., \mbox{$\RM(1,mm') \subset \code{2}{m} \subset \RM(2,mm')$}, and their generator matrix can be obtained from that of $\RM(2,mm')$ by removing a specific set of $m\binom{m'}{2}$ rows (Section~\ref{sec:subcodes_of_RM}). 
While $\code{2}{m}$ has the same minimum distance and a smaller rate than $\RM(2,mm')$, $\code{2}{m}$ has considerably fewer minimum weight codewords.
We identify the minimum weight codewords of $\code{2}{m}$ for all \mbox{$m,m' \geq 2$}, and show how its weight distribution can be computed for \mbox{$m'=2,3$} and all \mbox{$m \geq 2$} (Section~\ref{sec:min_weight_and_weight_dist}).
We tweak the existing soft-input decoders of second-order RM codes (belief propagation~\cite{LHP_ISIT_20} and local graph search~\cite{Kam_TCOM_22}) to decode $\code{2}{m}$ (Section~\ref{sec:decoding}). 
Simulations for lengths $256$, $512$ and $1024$ show that the CER versus $E_b/N_o$ performance of these RM subcodes is within $0.25$~dB of second-order RM codes and better than CRC-aided Polar codes (Section~\ref{sec:simulation_results}).
We conclude the paper in Section~\ref{sec:discussion}.

The RM subcodes $\code{2}{m}$ identified in this paper have rates close to $\sfrac{1}{2} \log_2^2 N$, which is higher than product codes whose component codes are first-order RM codes~\cite{JFMH_ICC_22} (rates close to $\sfrac{1}{4} \log_2^2 N$). 
For comparative rates, $\code{2}{m}$ have a smaller minimum distance than known algebraic subcodes of second-order RM codes~\cite[Chap.~15, Corollary~17 and Theorem~19]{macwilliams1977theory}, such as the Delsarte-Goethals codes. 
However, unlike these known subcodes, $\code{2}{m}$ can be decoded with minor modifications to existing decoders of second-order RM codes. 
This might be advantageous in terms of interoperability with RM codes in practical settings.

\emph{Notation:} The symbol $\otimes$ denotes the Kronecker product. For any positive integer $\ell$, let $[\ell] = \{1,\dots,\ell\}$.
The binary field is $\Fb_2=\{0,1\}$. 
The empty set is $\emptyset$.
For sets $A, B$ we define $A \setminus B = \left\{a \in A~:~ a \notin B \right\}$. 
We use capital letters to denote matrices (such as $G$), and small bold letters to denote row vectors (such as $\p{a}$).
The notation $(.)^T$ denotes the transpose operator.
The dimension of a linear code (subspace) $\mathcal{C}$ is denoted by $\dimenof(\mathcal{C})$, its minimum distance by $\dmin(\mathcal{C})$ and its dual code by $\mathcal{C}^\perp$. 
The support of a vector $\p{a}$ is denoted by $\supp(\p{a})$.
We use 
$\rowspaceof$ to denote the span of the rows of a matrix.

\section{Subcodes of Second-Order RM Codes via Recursive Subproducts} \label{sec:subcodes_of_RM}

In this section we identify a monomial basis for the recursive subproduct codes which use an RM code as a base code. 
When the base code is a first-order RM code $\Cs=\RM(1,m')$, we will see that the resulting second-order recursive subproduct code $\code{2}{m}$ is sandwiched between $\RM(1,mm')$ and $\RM(2,mm')$.

\subsection{Recursive Subproduct Codes with RM Base Code}

We first recall a recursive construction~\cite{SNK_ISIT24} of $\code{r}{m}$ for any $[n,k,d]$ base code $\Cs$ with $\allonelengthn \in \Cs$ and $k \geq 2$, where $\allonelengthn$ denotes the all-ones vector of length $n$.
Let $G_{sub}$ be any $(k-1) \times n$ matrix such that 
\begin{equation*}
G = \begin{bmatrix} \allonelengthn \\ G_{sub} \end{bmatrix}
\end{equation*}
is a generator matrix of $\Cs$. 
For \mbox{$0 < r < m$}, a generator matrix $G_{r,m}$ of $\code{r}{m}$ is recursively given by
\begin{equation} \label{eq:recursive_gen_matrix}
G_{r,m} = 
\begin{bmatrix}
G_{r,m-1} \otimes \allonelengthn \\
G_{r-1,m-1} \otimes  G_{sub}
\end{bmatrix}.
\end{equation}
When \mbox{$r=0$}, $\code{r}{m}$ is the repetition code of length $n^m$. For \mbox{$r=m$}, $\code{r}{m}$ is the $m$-fold product code $\Cs \otimes \cdots \otimes \Cs$, i.e., the generator matrix of $\code{m}{m}$ is $G^{\otimes m}$.

We want to consider the case where the base code is an RM code. 
Towards this, for any $\ell \geq 1$, we index the coordinates of vectors of length $2^\ell$ by the elements of $\Fb_2^\ell$ in the natural order~\cite[Chapter~13]{macwilliams1977theory}. That is, for $i \in [2^\ell]$, the $i^\tth$ entry of a vector is indexed by $(i_1,\dots,i_\ell) \in \Fb_2^\ell$ where $(i-1) = i_1 + 2 i_2 + 2^2 i_3 + \cdots + 2^{(\ell-1)}i_{\ell}$. For any function $f(x_1,\dots,x_\ell)$ in $\ell$ variables $x_1,\dots,x_\ell$, that is, $f:\Fb_2^{\ell} \to \Fb_2$, let $\eval(f) \in \Fb_2^{2^\ell}$ denote the vector obtained by evaluating $f$ at all points in $\Fb_2^\ell$. For $\p{\alpha} = (\alpha_1,\dots,\alpha_\ell) \in \Fb_2^\ell$, the $\p{\alpha}^\tth$ entry of $\eval(f)$ is $f(\alpha_1,\dots,\alpha_\ell)$. 
We use the following straightforward observation about the Kronecker product of the evaluations of two functions. 

\begin{lemma} \label{lem:kronecker_of_evals}
For any $\ell, \ell' \geq 1$ and functions $f: \Fb_2^{\ell} \to \Fb_2$ and $f':\Fb_2^{\ell'} \to \Fb_2$, the Kronecker product $\eval(f) \otimes \eval(f')$ equals the vector of length $2^{\ell + \ell'}$ obtained as the evaluation of 
\begin{equation*}
f'(x_1,x_2,\dots,x_{\ell'}) \times f(x_{\ell'+1},x_{\ell'+2},\dots,x_{\ell'+\ell}).
\end{equation*}
\end{lemma}
\begin{IEEEproof}
Note that $\eval(f) \otimes \eval(f')$ equals the block vector
$\begin{pmatrix}
f(\p{\alpha})\, \eval(f') \,:\, \p{\alpha} \in \Fb_2^{\ell}
\end{pmatrix}$, where the individual blocks $f(\p{\alpha})\, \eval(f')$ are placed in the natural order of the index $\p{\alpha} \in \Fb_2^{\ell}$.
The proof follows from the fact that the entries of $\eval(f) \otimes \eval(f')$ are indexed by the elements of $\Fb_2^{\ell + \ell'}$ in the natural order.
\end{IEEEproof}

We are now in a position to identify a basis (consisting of evaluations of monomials) for $\code{r}{m}$ when \mbox{$\Cs=\RM(r',m')$} for \mbox{$1 \leq r' \leq m'$}. 
This description involves $mm'$ variables $x_1,\dots,x_{mm'}$. For any \mbox{$S \subset [mm']$}, we define the monomial $x_S=\prod_{i \in S}x_i$, with $x_{\emptyset}=1$. 
Also, for given values of $m,m'$, we define $m$ sets 
\begin{equation} \label{eq:blocks_definition}
\begin{aligned}
\Bs_1 &= \{1,\dots,m'\}, \\ 
\Bs_2 &= \{m'+1,\dots,2m'\}, \\
&\vdots \\
\Bs_{m} &= \{mm'-m'+1,\dots,mm'\},
\end{aligned}
\end{equation}
and call these sets `blocks'. 

\begin{lemma} \label{lem:monomial_basis}
For any \mbox{$0 \leq r \leq m$} and \mbox{$1 \leq r' \leq m'$}, the following is a basis for the recursive subproduct code $\code{r}{m}$ when $\Cs=\RM(r',m')$
\begin{align*}
\Big\{ 
&\eval(x_{S_1 \cup \cdots \cup S_m})~:~S_i \subset \Bs_i, |S_i| \leq r' ~\forall~i \in [m], \text{ and}  \\
&~~~~\text{at the most } r \text{ sets among } S_1,\dots,S_m \text{ are non-empty}
\Big\}.
\end{align*}
\end{lemma}
\begin{IEEEproof}
We prove the lemma by showing that these monomial bases are the rows of a generator matrix of $\code{r}{m}$.
We choose the rows of $G_{sub}$ to be $\eval(x_S)$, where $S \subset [m]$ and $1 \leq |S| \leq r'$. This makes $G = [\allonelengthn^T~G_{sub}^T]^T$ a generator matrix of $\RM(r',m')$~\cite{macwilliams1977theory}.

We first provide a proof for all $0 < r < m$ using the recursion~\eqref{eq:recursive_gen_matrix}. The proof is by induction on the parameter $m$. When \mbox{$m=1$} we have $r \in \{0,1\}$. For $r=0$, $\code{r}{m}$ is the repetition code. The statement of the lemma holds in this case since the basis is $\{\eval(x_{\emptyset})\}=\{\nosubscriptallone\}$. 
For $r=m=1$, $\code{r}{m}=\Cs=\RM(r',m')$. The basis identified in the lemma is $\{\eval(x_S) \,: \, S \subset [m'], |S| \leq r'\}$, which is known to be a basis of $\RM(r',m')$.

We now prove the induction step using the recursion~\eqref{eq:recursive_gen_matrix} and Lemma~\ref{lem:kronecker_of_evals}. Assume that the lemma holds for the codes $\code{r-1}{m-1}$ and $\code{r}{m-1}$.
First consider those rows of $G_{r,m}$ that belong to the submatrix $G_{r,m-1} \otimes \allonelengthn$. 
Using the induction hypothesis, Lemma~\ref{lem:kronecker_of_evals} and the fact that $\allonelengthn=\eval(x_{\emptyset})$, we conclude that the rows of $G_{r,m-1} \otimes \allonelengthn$ are evaluations of $x_{\emptyset} \times x_{S_2 \cup \cdots \cup S_m} = x_{S_2 \cup \cdots \cup S_m}$ where $S_i \subset \Bs_i$, $|S_i| \leq r'$ and at the most $r$ sets among $S_2,\dots,S_m$ are non-empty. 
In other words, these are precisely the monomials $x_{S_1 \cup \cdots \cup S_m}$ identified in the present lemma for which $S_1 = \emptyset$. 
Now consider the rows in the submatrix $G_{r-1,m-1} \otimes  G_{sub}$. Any row here is the Kronecker product of one row each from $G_{r-1,m-1}$ and $G_{sub}$. Using induction, the rows of $G_{r-1,m-1}$ are $\eval(x_{S_1 \cup \cdots \cup S_{m-1}})$ where at the most $r-1$ of the sets $S_1,\dots,S_{m-1}$ are non-empty and $|S_1|,\dots,|S_{m-1}| \leq r'$. The rows from $G_{sub}$ are $\eval(x_S)$ with $1 \leq |S| \leq r'$. Using Lemma~\ref{lem:kronecker_of_evals}, we see that the Kronecker product is a basis $x_{S_1 \cup \cdots \cup S_m}$ identified in this lemma with $S_1$ being non-empty and at the most $r-1$ sets among $S_2,\dots,S_{m}$ being non-empty.
This completes the proof for all $0 < r < m$.

For $r=m$, a similar induction argument can be used based on the recursion $G_{m,m}=G_{m-1,m-1} \otimes G$. For $r=0$, the code $\code{r}{m}$ is a repetition code, and the proof of the lemma is straightforward.
\end{IEEEproof}

\begin{remark} \label{rem:translations}
Using Lemma~\ref{lem:monomial_basis} and the formalism of~\cite{BDOT_ISIT_2016}, we observe that recursive subproduct codes with RM base codes are \emph{weakly decreasing monomial codes}, i.e., if $x_S$ is a basis element and $S' \subset S$, then $x_{S'}$ is a basis element too. This implies that the automorphism group of these codes includes the translations of indices $\p{\alpha} \to \p{\alpha} + \p{\beta}$ for all $\p{\beta} \in \Fb_2^{mm'}$.
\end{remark}

\subsection{Subcodes of Second-Order RM Codes}

In this work we are interested in second-order recursive subproduct codes $\code{2}{m}$ with \mbox{$\Cs=\RM(1,m')$}, where $m,m' \geq 2$. Applying Lemma~\ref{lem:monomial_basis} to these codes we arrive at
\begin{corollary} \label{corr:monomial_basis_second_order}
For any $m,m' \geq 2$, the second-order recursive subproduct code $\code{2}{m}$ with base code $\Cs=\RM(1,m')$ has basis 
\begin{align*}
\big\{
\eval(x_S)~:~S \subset [mm'], |S| \leq 2, |S \cap \Bs_i| \leq 1~\forall~i \in [m]
\big\}.
\end{align*}
\end{corollary}

The monomial basis for $\RM(2,mm')$ consists of $x_{\emptyset}$, $x_1,\dots,x_{mm'}$ and all $x_ix_j$ where $i,j \in [mm']$ and $i \neq j$.
A basis for $\code{2}{m}$ can be obtained from the monomial basis for $\RM(2,mm')$ by removing all degree-$2$ monomials $x_ix_j$ such that $i$ and $j$ belong to the same block.

\begin{example}
Let the base code be $\Cs=\RM(1,2)$, which is the $[4,3,2]$ single-parity check code. 
Consider $\code{2}{m}$ with $m=3$. The blocks are 
\begin{equation*}
\Bs_1=\{1,2\}, ~\Bs_2=\{3,4\}, ~\Bs_3=\{5,6\}.
\end{equation*}
From Corollary~\ref{corr:monomial_basis_second_order}, the monomial basis for $\code{2}{3}$ consists of $x_{\emptyset}$, all the degree-$1$ terms $x_1,\dots,x_6$, and all degree-$2$ terms $x_ix_j$ except the following $3$ monomials: $x_1x_2$, $x_3x_4$, $x_5x_6$. These forbidden monomials are precisely the terms $x_ix_j$ where $i$ and $j$ belong to the same block. 
Removing the three rows $\eval(x_1x_2)$, $\eval(x_3x_4)$, $\eval(x_5x_6)$ from the generator matrix of $\RM(2,6)$ yields the generator matrix of $\code{2}{3}$.
\end{example}

\begin{example}
The generator matrix of $\RM(1,3)^{\otimes [2,3]}$ is obtained from the generator matrix of $\RM(2,9)$ by removing the following nine rows: $x_1x_2$, $x_1x_3$, $x_2x_3$, $x_4x_5$, $x_4x_6$, $x_5x_6$, $x_7x_8$, $x_7x_9$, $x_8x_9$. 
\end{example}

From Corollary~\ref{corr:monomial_basis_second_order}, observe that the basis monomials of $\RM(1,mm')$, i.e., $x_{\emptyset},x_1,x_2,\dots,x_{mm'}$, all belong to the basis of $\code{2}{m}$. Hence, $\RM(1,mm') \subset \code{2}{m}$. Also, every monomial $x_S$ in Corollary~\ref{corr:monomial_basis_second_order} satisfies $|S| \leq 2$. Hence, $\code{2}{m} \subset \RM(2,mm')$. 
Thus, we have

\begin{corollary} \label{corr:sandwich_second_order}
For any $m,m' \geq 2$ and $\Cs=\RM(1,m')$ we have $\RM(1,mm') \subset \code{2}{m} \subset \RM(2,mm')$.
\end{corollary}

The dimension of $\code{2}{m}$ is the number of elements in the basis which equals 
\begin{equation*}
1 + mm' + \binom{m}{2} {m'}^2.
\end{equation*}
From~\eqref{eq:basic_parameters_recursive_subproduct_codes}, we see that $\dmin(\code{2}{m}) = 2^{mm'-2}$, which equals the minimum distance of $\RM(2,mm')$.
Thus, $\code{2}{m}$ has the same length and minimum distance as $\RM(2,mm')$, but with a smaller dimension than $\RM(2,mm')$. 
This smaller code dimension leads to a considerable reduction in the number of minimum weight codewords, which is a code parameter (along with $\dmin$) that determines the CER at high signal-to-noise ratio.

\section{Minimum Weight Codewords \& Weight Distribution} \label{sec:min_weight_and_weight_dist}

Unless otherwise stated, in the rest of the paper we will assume that $\Cs=\RM(1,m')$ for some $m' \geq 2$.
In this section we identify the minimum weight codewords of $\code{2}{m}$ for all $m,m' \geq 2$. 
Then, using the well known framework~\cite[Chapter~15]{macwilliams1977theory} that relates the weight distribution of a coset of first-order RM code (in second-order RM code) to the rank of a symplectic matrix, we identify a recursion to numerically determine the weight distribution of $\code{2}{m}$ for $m'=2,3$ and all $m \geq 2$.

\subsection{Minimum Weight Codewords}

For any linear code $\mathcal{C}$, let $\Amin(\mathcal{C})$ denote the collection of codewords in $\Cs$ with Hamming weight equal to $\dmin(\mathcal{C})$.
Since $\code{2}{m} \subset \RM(2,mm')$ and since both these codes have the same minimum distance, $\Amin(\code{2}{m}) \subset \Amin(\RM(2,mm'))$. 
Thus, $\Amin(\code{2}{m})$ is the collection of vectors in $\Amin(\RM(2,mm'))$ that lie in the span of the monomial basis identified in Corollary~\ref{corr:monomial_basis_second_order}.

We first recall the characterization~\cite{macwilliams1977theory} of the minimum weight codewords of $\RM(2,mm')$, viz., $\p{c} \in \Amin\left( \RM(2,mm') \right)$ if and only if $\supp(\p{c})$ is a coset of an $(mm'-2)$-dimensional subspace of $\Fb_2^{mm'}$.
Every such coset can be described using two linearly independent vectors $\p{a}_1,\p{a}_2 \in \Fb_2^{mm'}$, and two scalars $b_1,b_2 \in \Fb_2$. 
The span of $\{\p{a}_1,\p{a}_2\}$ is the subspace orthogonal to the $(mm'-2)$-dimensional subspace of which $\supp(\p{c})$ is a coset.
The support of the codeword $\p{c} \in \Amin(\RM(2,mm'))$ corresponding to $(\p{a}_1,\p{a}_2,b_1,b_2)$ is 
\begin{equation} \label{eq:coset_description_Ax_equals_b}
\supp(\p{c}) = \left\{ \p{\alpha} \in \Fb_2^{mm'}~:~ \p{a}_1\p{\alpha}^T=b_1 \text{ and } \p{a}_2\p{\alpha}^T = b_2 \right\}.
\end{equation}
Further, $\p{c}$ is the evaluation of the polynomial
\begin{equation} \label{eq:poly_description_min_weight_codeword}
\left( \sum_{i=1}^{mm'} a_{1,i} x_i + b_1 + 1 \right) \left( \sum_{j=1}^{mm'}a_{2,j}x_j + b_2 + 1\right),
\end{equation}
where $\p{a}_1=(a_{1,1},\dots,a_{1,mm'})$ and $\p{a}_2=(a_{2,1},\dots,a_{2,mm'})$.
Note that for every coset, there are six possible choices of $(\p{a}_1,\p{a}_2,b_1,b_2)$ such that~\eqref{eq:coset_description_Ax_equals_b} equals this coset; this is a consequence of the fact that there are six distinct ordered bases for any $2$-dimensional subspace over $\Fb_2$.

Consider the $2 \times mm'$ matrix $A$ whose two rows are $\p{a}_1,\p{a}_2$. Since these two rows are linearly independent $\rank(A)=2$. The coefficient of the term $x_ix_j$ in the polynomial~\eqref{eq:poly_description_min_weight_codeword} is 
\begin{equation*}
\det 
\begin{bmatrix}
a_{1,i} & a_{1,j} \\ a_{2,i} & a_{2,j}
\end{bmatrix},
\end{equation*}
which is the determinant of the $2 \times 2$ submatrix of $A$ formed by its $i^\tth$ and $j^\tth$ columns.
Now, using Corollary~\ref{corr:monomial_basis_second_order} we see that the codeword $\p{c} \in \Amin( \RM(2,mm') )$ corresponding to~\eqref{eq:poly_description_min_weight_codeword} lies in $\code{2}{m}$ if and only if the above determinant is zero for all choices of $i,j \in [mm']$ such that $i$ and $j$ belong to the same block.
This condition is equivalent to requiring that for each of the blocks $\Bs_i$, $i \in [m]$, the submatrix of $A$ formed by the columns indexed by $\Bs_i$ has rank at the most $1$.
Using the definition of the blocks~\eqref{eq:blocks_definition}, this condition is equivalent to the following: writing $A$ as the block matrix 
\begin{equation} \label{eq:matrix_A_for_second_order_codewords}
A 
= \begin{bmatrix} \p{a}_1 \\ \p{a}_2  \end{bmatrix}
= \begin{bmatrix} A_1 & A_2 & \cdots & A_m \end{bmatrix}  
\end{equation}
where $A_1,\dots,A_m$ are $2 \times m'$ matrices, we require $\rank(A)=2$ and $\rank(A_1),\dots,\rank(A_m) \leq 1$. 
Observe that for any such $A$, i.e., for any such choice of $(\p{a}_1,\p{a}_2)$, there are four choices of $(b_1,b_2)$ so that $(\p{a}_1,\p{a}_2,b_1,b_2)$ corresponds to a codeword in $\Amin( \code{2}{m} )$ via~\eqref{eq:coset_description_Ax_equals_b}. 
Using this observation along with the fact that the map $(\p{a}_1,\p{a}_2,b_1,b_2) \to \p{c}$ is a six-to-one map, we arrive at 
\begin{lemma} \label{lem:num_of_min_wt_using_A}
The number of minimum weight codewords in $\code{2}{m}$, for $\Cs=\RM(1,m')$, is 
\begin{equation*}
\frac{4}{6} \times 
\left| \, 
\big\{ 
A~:~\rank(A)=2, \, \rank(A_1),\dots,\rank(A_m) \leq 1
\big\}
\, \right|.
\end{equation*}
\end{lemma}

Our next result identifies the number of matrices $A$ that satisfy the constraints on the rank imposed by Lemma~\ref{lem:num_of_min_wt_using_A}.
\begin{lemma} \label{lem:counting_A_matrices}
The number of matrices $A$ such that $\rank(A)=2$ and $\rank(A_i) \leq 1$ for all $i \in [m]$ is 
\begin{equation*}
(3 \times 2^{m'} - 2)^m - (3 \times 2^{m m'} - 2).   
\end{equation*}
\end{lemma}
\begin{IEEEproof}
We first count the number of matrices $A$ satisfying the $m$ constraints $\rank(A_i) \leq 1$, $i \in [\ell]$, and then remove the cases with $\rank(A) \leq 1$. We invoke the property that the rank of any submatrix $A_i$ is not greater than the rank of the matrix $A$. If $\rank(A) \leq 1$, then $\rank(A_{i}) \leq 1$ for all $i$, thereby making the removal step feasible.

For each $i \in [m]$, consider the number of $2 \times m'$ matrices $A_i$ with $\rank(A_{i}) \leq 1$. Rank of $A_i$ is $1$ if and only if 
\emph{(i)}~both the rows of $A_i$ are identical and non-zero, or 
\emph{(ii)}~the first row of $A_i$ is non-zero and the second row is zero, or
\emph{(iii)}~the second row of $A_i$ is non-zero and the first row is zero.
Also, there is exactly one matrix with rank $0$. Hence, the number of $2 \times m'$ matrices with rank at the most $1$ is 
$3 \times (2^{m'} - 1) + 1 = 3 \times 2^{m'} - 2$.

Note that, if we do not impose the condition $\rank(A) = 2$, the submatrices $A_{i}$ and $A_j$ can be chosen independently for all $i \neq j$. Therefore, the number of matrices $A$ with $\rank(A_i) \leq 1$ for all $i \in [m]$ is $(3 \times 2^{m'} - 2)^{m}$.

To impose the condition $\rank(A) = 2$, we subtract the cases where $\rank(A) \leq 1$. 
The arguments used above for counting the number of $2 \times m'$ matrices $A_i$ with rank at the most $1$ can be used again. 
The only difference is the number of columns---$A$ has $mm'$ columns while $A_i$ has $m'$ columns.
Therefore, $3 \times 2^{mm'} - 2$ matrices correspond to the case $\rank(A) \leq 1$.

Putting everything together, the number of matrices satisfying all the constraints in the lemma is 
\begin{equation*}
(3 \times 2^{m'} - 2)^{m} - (3 \times 2^{mm'} - 2). 
\end{equation*}
\end{IEEEproof}

The main result of this subsection follows immediately from Lemmas~\ref{lem:num_of_min_wt_using_A} and~\ref{lem:counting_A_matrices}.

\begin{theorem}
For \mbox{$m,m' \geq 2$}, and \mbox{$\Cs=\RM(1,m')$}, the number of minimum weight codewords in $\code{2}{m}$ is 
\begin{equation*}
\frac{2}{3} \left( (3 \times 2^{m'} - 2)^{m} - 3 \times 2^{mm'} + 2 \right).
\end{equation*}
\end{theorem}

\subsubsection*{Comparison with Second-Order RM Codes}

Using $N$ to denote the block length of the code, observe that the dimension and the number of minimum weight codewords of second-order RM codes are~\cite[Chapter~13, Theorem~9]{macwilliams1977theory} 
\begin{align*}
\dim \left( \RM(2,mm') \right) &= \frac{1}{2} \log_2^2 N + \frac{1}{2} \log_2 N + 1, \\
\left| \Amin \left( \RM(2,mm') \right) \right| &= \frac{2}{3} \left( N^2 - 3N + 2 \right).
\end{align*}
In comparison, $\code{2}{m}$ has
\begin{align*}
\dim \left( \code{2}{m} \right) &= \frac{1}{2} \log_2^2 N - \frac{m'-2}{2} \log_2 N + 1, \\
\left| \Amin \left( \code{2}{m} \right) \right| &= \frac{2}{3} \left( N^\gamma - 3N + 2 \right),
\end{align*}
where \mbox{$\gamma = ( \log_2 ( 3 \times 2^{m'} - 2 ) ) / m' < 2$} for all $m' \geq 2$. 
Thus, while the dimensions of both the codes grow as $\sim \frac{1}{2} \log_2 ^2 N$, the number of minimum weight codewords of $\code{2}{m}$ grows slower than the RM codes, such as $\Theta(N^{1.66})$ (for $m'=2$) versus $\Theta(N^2)$.

\begin{example}
We compare $\RM(2,8)$ with $\code{2}{4}$ where \mbox{$\Cs=\RM(1,2)$}. Both codes have length $256$ and minimum distance $64$. 
The dimensions of $\RM(2,8)$ and $\code{2}{4}$ are $37$ and $33$, respectively. The number of minimum weight codewords for these two codes are $43180$ and $6156$, respectively.
\end{example}

\begin{remark} \label{rem:optimality_of_RM_subcodes}
The generator matrix for $\code{2}{m}$ is obtained from the generator matrix of $\RM(2,mm')$ by removing a specific set of $\binom{m'}{2}m$ rows corresponding to degree-$2$ monomials. The generator matrix of $\RM(2,mm')$ contains $\binom{mm'}{2}$ rows of degree-$2$, which is larger than $\binom{m'}{2}m$. 
In general, if we want a subcode of $\RM(2,mm')$ with the same dimension as $\code{2}{m}$, we do not know which choice of $\binom{m'}{2}m$ degree-$2$ rows (to be removed) yields a code with the least number of minimum weight vectors. However, for the cases \mbox{$m'=2$} and \mbox{$m=2,3,4$}, we numerically verified that removing the rows $x_i x_j$, where $i$ and $j$ belong to the same block, yields the code with the smallest number of minimum weight codewords. Hence, the codes $\RM(1,2)^{\otimes[2,2]}$, $\RM(1,2)^{\otimes[2,3]}$ and $\RM(1,2)^{\otimes[2,4]}$ are optimal from this point of view.
We believe that this optimality extends to other values of $m,m'$ too, but we do not have a proof.
\end{remark}

\subsection{Weight Distribution}

The weight distribution of a code (or a coset of a code) of length $N$ is the list $\wtdist_0,\wtdist_1,\dots,\wtdist_N$, where $\wtdist_w$ is the number of vectors of weight $w$ in the code (or the coset). 
In this subsection, we numerically identify the weight distribution of $\code{2}{m}$ for all \mbox{$m \geq 2$}, when $\Cs=\RM(1,2)$ and $\Cs=\RM(1,3)$. 
We do not have a general result that applies to all $\Cs=\RM(1,m')$ with $m' \geq 2$.

\subsubsection{Weight Distribution of Cosets of $\RM(1,mm')$~\cite{macwilliams1977theory}}

We first recall from~\cite{macwilliams1977theory} how the weight distributions of cosets of first-order RM codes in second-order RM codes can be determined via the rank of certain binary \emph{symplectic} matrices (i.e., binary symmetric matrices with all-zero diagonal). 
The coset of $\RM(1,mm')$ in $\RM(2,mm')$ that contains the codeword 
\begin{equation*}
\eval \left( \, \sum_{i=1}^{mm'-1}\sum_{j=i+1}^{mm'} b_{i,j} x_ix_j \, \right)
\end{equation*}
is associated with the $mm' \times mm'$ symplectic matrix $B$ whose $(i,j)$ and $(j,i)$ entries are equal to $b_{i,j}$ for all $i<j$.
It is known that the rank of any symplectic matrix is even. If $\rank(B)=h$, where $h$ is even, then 
Table~\ref{table:weight_dist_of_coset_of_RM_first_order} gives the weight distribution of the coset of $\RM(1,mm')$ corresponding to $B$~\cite[Chapter~15, Theorem~5]{macwilliams1977theory}. 
\begin{table} 
\centering
\renewcommand{\arraystretch}{1.5}
\caption{Weight distribution of the coset of $\RM(1,mm')$ in $\RM(2,mm')$ corresponding to a symplectic matrix of rank $h$. Value $\wtdist_w=0$ for all other choices of $w$.}
\begin{tabular}{c|c} 
\hline
\hline  
Weight $w$ & Number of Codewords $\wtdist_w$ \\
\hline 
$2^{mm'-1} - 2^{mm'-1-\sfrac{h}{2}}$ & $2^h$ \\
\hline 
$2^{mm'-1}$ & $2^{mm'+1} - 2^{h+1}$ \\
\hline 
$2^{mm'-1} + 2^{mm'-1-\sfrac{h}{2}}$ & $2^h$ \\
\hline 
\end{tabular}
\label{table:weight_dist_of_coset_of_RM_first_order}
\end{table}

\subsubsection{Weight Distribution of $\code{2}{m}$} \label{sec:sub:weight_distribution_new_codes}

Since 
\begin{equation*}
\RM(1,mm') \subset \code{2}{m} \subset \RM(2,mm'),
\end{equation*}
every coset of $\RM(1,mm')$ in $\code{2}{m}$ is also a coset of $\RM(1,mm')$ in $\RM(2,mm')$. 
Each coset of $\RM(1,mm')$ in $\code{2}{m}$ can be identified using a representative codeword 
\begin{equation*}
\eval \left( \, \sum_{i=1}^{mm'-1}\sum_{j=i+1}^{mm'} b_{i,j} x_ix_j \, \right)
\end{equation*}
where $b_{i,j}=0$ if $i$ and $j$ belong to the same block. This is a consequence of Corollary~\ref{corr:monomial_basis_second_order}. Thus, the symplectic matrix corresponding to any coset will be of the form 
\begin{equation} \label{eq:B_block_matrix}
B = 
\begin{bmatrix}
0 & B_{1,2} & \cdots & B_{1,m} \\
B_{2,1} & 0 & \cdots & B_{2,m} \\ 
\vdots  &  \vdots & \ddots & \vdots \\
B_{m,1} & B_{m,2} & \cdots & 0
\end{bmatrix},
\end{equation}
where each $B_{i,j}$ is an $m' \times m'$ matrix, $B_{j,i} = B_{i,j}^T$ and all the diagonal blocks $B_{i,i}=0$. 
We will denote the collection of all such matrices by $\mathcal{B}_{m,m'}$, or simply by $\mathcal{B}$ if the values $m,m'$ are clear from the context. That is,
\begin{equation*}
\mathcal{B}_{m,m'} = \left\{ B \in \Fb_2^{mm' \times mm'}  : B^T  = B, B_{i,i} =0 ~\forall~ i \in [m] \right\}.
\end{equation*}
Note that $\mathcal{B}_{m,m'}$ is closed under addition and 
\begin{align*}
\log_2 |\mathcal{B}_{m,m'}| &= \binom{m}{2}m'^2 \\
&= \dim\left(\code{2}{m}\right) - \dim\left(\RM(1,mm')\right).
\end{align*}
The cosets of $\RM(1,mm')$ in $\code{2}{m}$ are in a one-to-one correspondence with the matrices in $\mathcal{B}$. The weight distribution of a coset corresponding to $B \in \mathcal{B}$ with $\rank(B)=h$ is given by Table~\ref{table:weight_dist_of_coset_of_RM_first_order}. 
Since the weight distribution of $\code{2}{m}$ is the sum of the weight distributions of all the cosets, in order to numerically compute the weight distribution of $\code{2}{m}$ it is sufficient to determine the number of matrices in $\mathcal{B}$ with rank $h$ for $h \in \{0,1,\dots,mm'\}$. We will denote this as 
\begin{equation*}
N_{m'}(m,h) = \left| \, \left\{ B \in \mathcal{B}_{m,m'}~:~\rank(B)=h \right\} \, \right|.
\end{equation*}

We first consider the case $m'=2$, and provide a recurrence relation for $N_{2}(m,h)$ which can be immediately used to compute the weight distribution of $\code{2}{m} = \RM(1,2)^{\otimes [2,m]}$ for all $m \geq 2$ (we do not have a closed-form expression for $N_2(m,h)$). Note that the derivation of weight distribution of second-order RM codes in~\cite{macwilliams1977theory} uses a recurrence as well. 

Observe that $N_{2}(m,h)=1$ for $h=0$ and any $m \geq 1$, since the only matrix in $\mathcal{B}$ with rank $0$ is the all-zero matrix. 
If \mbox{$m = 1$} and $h \in \{1,2\}$, then $N_{2}(m,h) = 0$, since in this case $\mathcal{B} = \{0\}$. 
If $m < 1$, $h < 0$ or $h > 2m$, then we have $N_{2}(m,h) = 0$.

\begin{lemma} \label{lem:recurrence_mdash_2}
For any $h \in [2m]$, $N_2(m,h)$ equals 
\begin{align*}
t_1 N_2(m \!-\! 1,h) &+ t_2 N_2(m \!-\! 1,h\!-\!2) + t_3N_2(m \! - \!1,h \! - \! 4) \\
\text{where } t_1 &= 2^{h-1} + 2^{2h-1}, \\
t_2 &= 3 \times 2^{2m - 4 + h} - 5 \times 2^{2h - 5} - 2^{h - 3},\\ 
t_3 &= -3 \times 2^{2m - 6 + h} + 4^{2m - 2} + 2^{2h - 7}.
\end{align*}
\end{lemma}
\begin{IEEEproof}
The proof is available in Appendix~\ref{app:lem:recurrence_mdash_2}.
\end{IEEEproof}

We now consider the case \mbox{$m' = 3$} and provide a recurrence relation for $N_{3}(m,h)$. The recursion boundary conditions are similar to the $m' = 2$ case, viz.,
$N_{3}(m, h) = 0$ if $m < 1, h < 0$, or $h > 3m$; $N_{3}(m, h)=0$ if $m = 1$ and $h \in \{1, 2, 3\}$; and $N_{3}(m, h) = 1$ for all $m \geq 1$ and $h = 0$. 

\begin{lemma} \label{lem:recurrence_mdash_3}
For any $h \in [3m]$, $N_3(m, h)$ equals 
\begin{align*}
    t_1 N_3(m - 1, h) + t_2 N_3(m - 1, h - 2) 
 + t_3 N_3(m - 1, h - 4) + t_4 N_3(m - 1, h - 6),
\end{align*}
where
\begin{align*}
    t_1 &= 2^{3h - 3} + 7 \times 2^{2h - 3}, \\
    t_2 &= (2^{3m - 3} - 2^{h - 2}) \times (2^{2h - 3} + 2^{h} + 1) + (2^{3m - 3} - 2^{h - 3}) \times (2^{2h - 4} + 2^{h - 1} - 3) \\ &~~~~+ (2^{3m - 3} - 2^{h - 4}) \times (2^{2h - 5} - 5\times 2^{h - 3} + 2), \\
    t_3 &= (2^{3m - 3} - 2^{h - 4}) \times (7 \times 2^{3m + h - 7} -21 \times 2^{2h - 9} -7 \times 2^{h - 5}), \\
    t_4 &= \prod_{i=4}^{6} (2^{3m - 3} - 2^{r - i}).
\end{align*}
\end{lemma}

\begin{IEEEproof}
The proof is similar to the proof of Lemma~\ref{lem:recurrence_mdash_2}. We provide a part of the proof (derivation of $t_1$) in Appendix~\ref{app:lem:recurrence_mdash_3} to illustrate the ideas involved.
\end{IEEEproof}

\section{Decoding} \label{sec:decoding}

In this section we adapt the following decoding algorithms of second-order RM codes to decode $\code{2}{m}$ in the binary-input additive white Gaussian noise (AWGN) channel: 
\emph{(i)}~the belief-propagation (BP) decoder from~\cite{LHP_ISIT_20} (which is based on recursive projection aggregation~\cite{YeA_IT_20}); and 
\emph{(ii)}~the local graph search (LGS) decoder~\cite{Kam_TCOM_22}. 
We use the LGS decoder to attain improvements over the CER performance of the BP decoder.

\subsection{Belief-Propagation Decoding}

The BP decoder for second-order RM codes from~\cite{LHP_ISIT_20} uses $N-1$ generalized check nodes in the Tanner graph, where each such node corresponds to one possible projection operation~\cite{YeA_IT_20}. These nodes in~\cite{LHP_ISIT_20} employ a fast soft-in soft out decoder (SISO) for first-order RM codes.

In the case of $\code{2}{m}$, some of these codeword projections belong to $\RM(1,d) \otimes \RM(0,mm'-d-1)$ (for some integer $d$), which is a proper subcode of first-order RM code $\RM(1,mm'-1)$ when $d < mm'-1$; for these projections we use fast SISO decoders of these corresponding subcodes. 
Further, the codewords of $\code{2}{m} = \RM(1,m')^{\otimes [2,m]}$ satisfy more constraints than the codewords of $\RM(2,mm')$; in particular, $\RM(1,m')^{\otimes [2,m]}$ is a subcode of the $m$-fold product code $\RM(1,m')^{\otimes m}$. Our decoder uses additional generalized check nodes (that perform SISO decoding of $\RM(1,m')$) to exploit this structure. 
We describe these modifications next.

\subsubsection{Projections} 

Consider a codeword of $\RM(2,mm')$ which is an evaluation of the polynomial 
\begin{align*}
f(\p{x}) = b_0 + \sum_{i=1}^{mm'} b_i x_i + \sum_{i=1}^{mm'-1}\sum_{j=i+1}^{mm'} b_{i,j} x_i x_j,     
\end{align*}
where \mbox{$b_i,b_{i,j} \in \Fb_2$}. 
The sum of this codeword and its permutation by a translate $\p{a} \in \Fb_2^{mm'} \setminus \{ \p{0} \}$ is the evaluation of 
\begin{align*}
f(\p{x}) + f(\p{x} + \p{a}) = \sum_{i=1}^{mm'} b_i a_i + \sum_{i < j}b_{i,j}a_ia_j + \p{a}B\p{x}^T,
\end{align*}
where $B$ is the symplectic matrix with $b_{i,j}$ as the $(i,j)$ entry for all $i<j$. 
For any codeword in $\code{2}{m}$, the matrix $B$ belongs to the set $\mathcal{B}_{m,m'}$ identified in Section~\ref{sec:sub:weight_distribution_new_codes}. 
Consider 
\begin{equation*}
U_{\p{a}} = \left\{ \p{a}B~:~B \in \mathcal{B}_{m,m'} \right\}, 
\end{equation*}
which is a subspace of $\Fb_2^{mm'}$ (since $\mathcal{B}_{m,m'}$ is closed under addition). 
Note that the sum of any codeword $f(\p{x})$ and its permutation $f(\p{x} + \p{a})$ belongs to
\begin{align*}
\left\{ \,
\eval\left( \, u_0 + \p{u}\p{x}^T \, \right)~:~u_0 \in \Fb_2, \, \p{u} \in U_{\p{a}}
\, \right\}.
\end{align*}
This is a subcode of $\RM(1,mm')$. 
It is easy to see that under an appropriate permutation of coordinates ($\p{x} \to A\p{x}$, for a suitable choice of matrix $A$ with $\det(A) \neq 0$) this code is transformed to 
\begin{align*}
\left\{ \,
\eval\left( \, u_0 + \sum_{i=1}^{d} u_i x_i \, \right)~:~u_0,\dots,u_d \in \Fb_2
\, \right\},
\end{align*}
where $d=\dim(U_{\p{a}})$ and the evaluation is done at all points $\p{\alpha} \in \Fb_2^{mm'}$. 
Observe that the evaluation at $\p{\alpha}$ is independent of the last $mm'-d$ entries of $\p{\alpha}$. 
Hence, this code is the product of $\RM(1,d)$ with the repetition code of length $2^{mm'-d}$, i.e., $\RM(1,d) \otimes \RM(0,mm'-d)$.

While computing the log-likelihood ratios (LLRs) for $\eval \left( f(\p{x}) + f(\p{x} + \p{a}) \right)$ from the channel observations a `box-plus' operation $\boxplus$ is used, see~\cite{IvU_ITW_2019} for instance. 
For any $\p{\alpha}$, the computed LLRs for the coordinates $\p{\alpha}$ and $\p{\alpha} + \p{a}$ are identical. 
This uninformative repetition is discarded by retaining only half the computed LLRs. Hence, the corresponding projected codebook is $\RM(1,d) \otimes \RM(0,mm'-d-1)$, which is of length $2^{mm'-1}$. We summarize this as 

\begin{lemma}
Up to a permutation of coordinates, the projection of $\code{2}{m}$ obtained using the translation vector $\p{a} \in \Fb_2^{mm'}$ is $\RM(1,d) \otimes \RM(0,mm'-d-1)$, where $d = \dim(U_{\p{a}})$.
\end{lemma}

Note that this product code can be SISO decoded using straightforward modifications of any available SISO decoder of first-order RM codes. For numerical stability, in our simulations we use the efficient max-log-MAP decoder for $\RM(1,d)$ proposed in~\cite{SNK_ISIT24}, which operates in the log domain (unlike the MAP decoder of~\cite{AsL_IT_04} that operates in the probability domain) and has complexity $\mathcal{O}(d \, 2^d)$. 
Our Tanner graph uses $N-1$ projections, one for each non-zero $\p{a}$, and a SISO max-log-MAP decoder for each of these projections.

Our next result identifies the dimension $d=\dim(U_{\p{a}})$ for every non-zero $\p{a}$.
\begin{lemma} \label{lem:dimension_of_Ua}
For any non-zero $\p{a} \in \Fb_2^{mm'}$ we have $\dim(U_{\p{a}}) = m'(m-1)$ if $\supp(\p{a})$ is a subset of one of the blocks $\Bs_1,\dots,\Bs_m$. Otherwise, $\dim(U_{\p{a}}) = mm' - 1$.
\end{lemma}
\begin{IEEEproof}
Please see Appendix~\ref{app:lem:dimension_of_Ua}.
\end{IEEEproof}


\subsubsection{Generalized Check Nodes of Product Codes}

All the codewords of $\code{2}{m}$ belong to the $m$-fold product code $\RM(1,m') \otimes \cdots \otimes \RM(1,m')$. Hence, if a codeword $\p{c}$ is viewed as an $m$-dimensional array (of length $2^{m'}$ in each dimension), any length-$2^{m'}$ sub-vector of $\p{c}$ oriented along any of the $m$ directions must belong to $\RM(1,m')$. 
Our BP decoder uses one generalized check node for each such sub-vector (see~\cite[Fig.~3]{Tan_IT_81}). These nodes perform max-log-MAP decoding of $\RM(1,m')$~\cite[Sec.~III-B]{SNK_ISIT24} during BP iterations.
The number of such generalized check nodes is $m2^{m'(m-1)}$.

\subsection{Local Graph Search Decoding}

The LGS decoder~\cite{Kam_TCOM_22} traces a `path' (of some specified length, say $P_{\sf LGS}$) of codewords such that the Hamming distance between consecutive codewords in this path equals the minimum distance $\dmin$ of the code, and no two codewords in this path are identical. The initial codeword $\p{c}^{(0)}$ is a suboptimal estimate; in our simulations, this is the codeword determined by the BP decoder. For $i \in [P_{\sf LGS}]$, the $i^\tth$ codeword $\p{c}^{(i)}$ is to be chosen from among all codewords (that have not yet been visited) at a distance $\dmin$ from $\p{c}^{(i-1)}$, i.e., from the neighbors of $\p{c}^{(i-1)}$ in the code space. 
Ideally, the neighbor of $\p{c}^{(i-1)}$ that has the largest maximum-likelihood (ML) metric (i.e., the shortest Euclidean distance from the channel output) is to be selected as $\p{c}^{(i)}$.
Finally, from among all the codewords in the path $\p{c}^{(0)},\dots,\p{c}^{(P_{\sf LGS})}$, the decoder outputs the codeword with the largest ML metric.

For decoding second-order RM codes $\RM(2,mm')$,~\cite{Kam_TCOM_22} uses a suboptimal greedy algorithm to identify each $\p{c}^{(i)}$, $i \in [P_{\sf LGS}]$. 
In this technique, given $\p{c}^{(i-1)}$, the support $\supp\left( \p{c}^{(i)} - \p{c}^{(i-1)} \right)$ (which is a coset of an $(mm'-2)$-dimensional subspace in $\Fb_2^{mm'}$) is identified in two steps: first identify a coset of an $(mm'-1)$-dimensional subspace, and then identify $\supp(\p{c}^{(i)}- \p{c}^{(i-1)})$ as a subset of this coset.
Viewing $\p{c}^{(i)} - \p{c}^{(i-1)}$ as the evaluation of the product polynomial~\eqref{eq:poly_description_min_weight_codeword}, this two-step process is identical to determining the two factors of the product, viz., $\sum_{i=1}^{mm'}a_{1,i}x_i + b_1+1$ and $\sum_{i=1}^{mm'}a_{2,i}x_i + b_2+1$, one after the other.

In order to adapt this decoder for $\code{2}{m}$, we use the first step (that of determining $\sum_{i=1}^{mm'}a_{1,i}x_i + b_1+1$) as in~\cite{Kam_TCOM_22} except that we do not allow $a_{1,1}=\cdots=a_{1,mm'}=0$. For the second step, we apply straightforward modifications to the algorithm of~\cite{Kam_TCOM_22} in order to consider only those polynomials $\sum_{i=1}^{mm'}a_{2,i}x_i + b_2+1$ such that the corresponding matrix $A$ in~\eqref{eq:matrix_A_for_second_order_codewords} satisfies $\rank(A)=2$ and $\rank(A_i) \leq 1$ for all $i \in [m]$.
This ensures that $\p{c}^{(i)}- \p{c}^{(i-1)} \in \Amin\left(\code{2}{m}\right)$, and consequently, $\p{c}^{(i)} \in \code{2}{m}$ for all $i \in [P_{\sf LGS}]$.

\begin{figure}[!t]
    \centering
    \includegraphics[width=0.75\columnwidth]{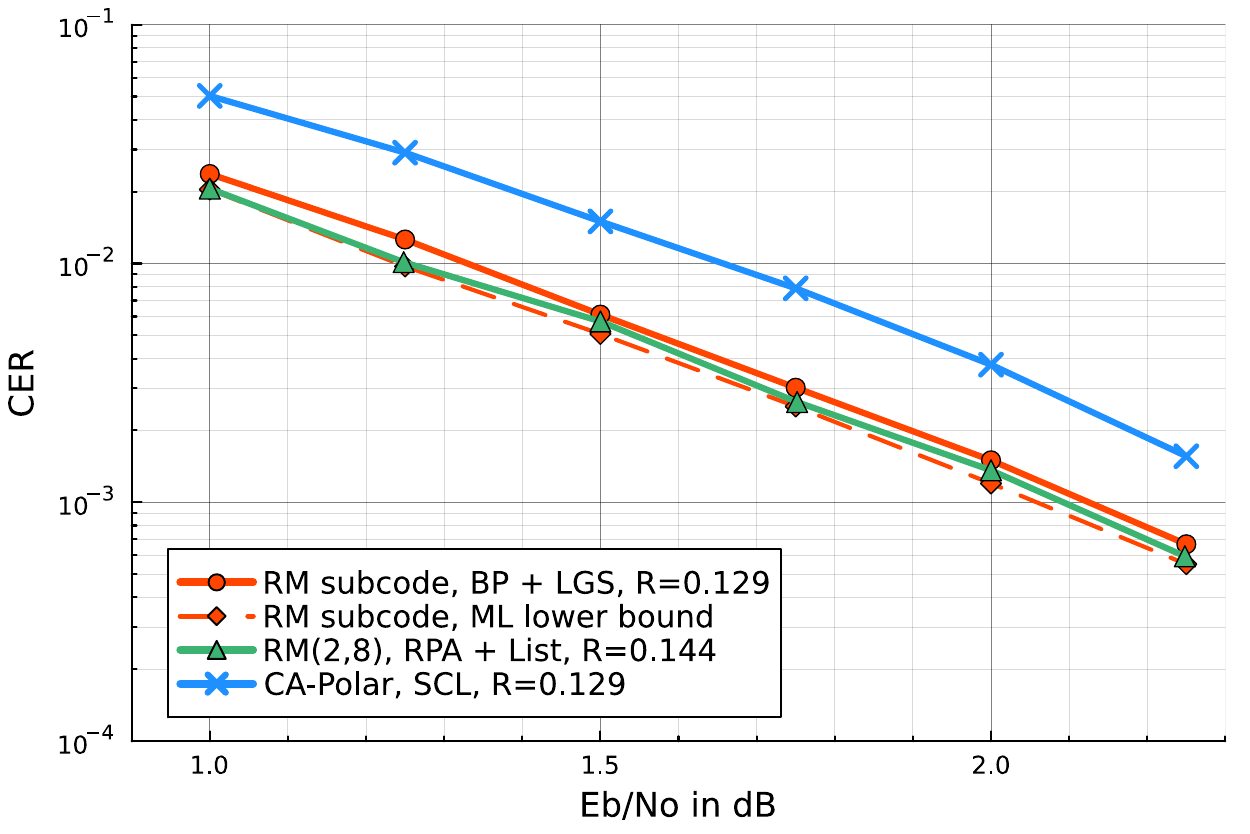}
    \caption{Codeword error rates of codes with length $256$.}
    \label{fig:256}
\end{figure}
\begin{figure}[!t]
    \centering
    \includegraphics[width=0.75\columnwidth]{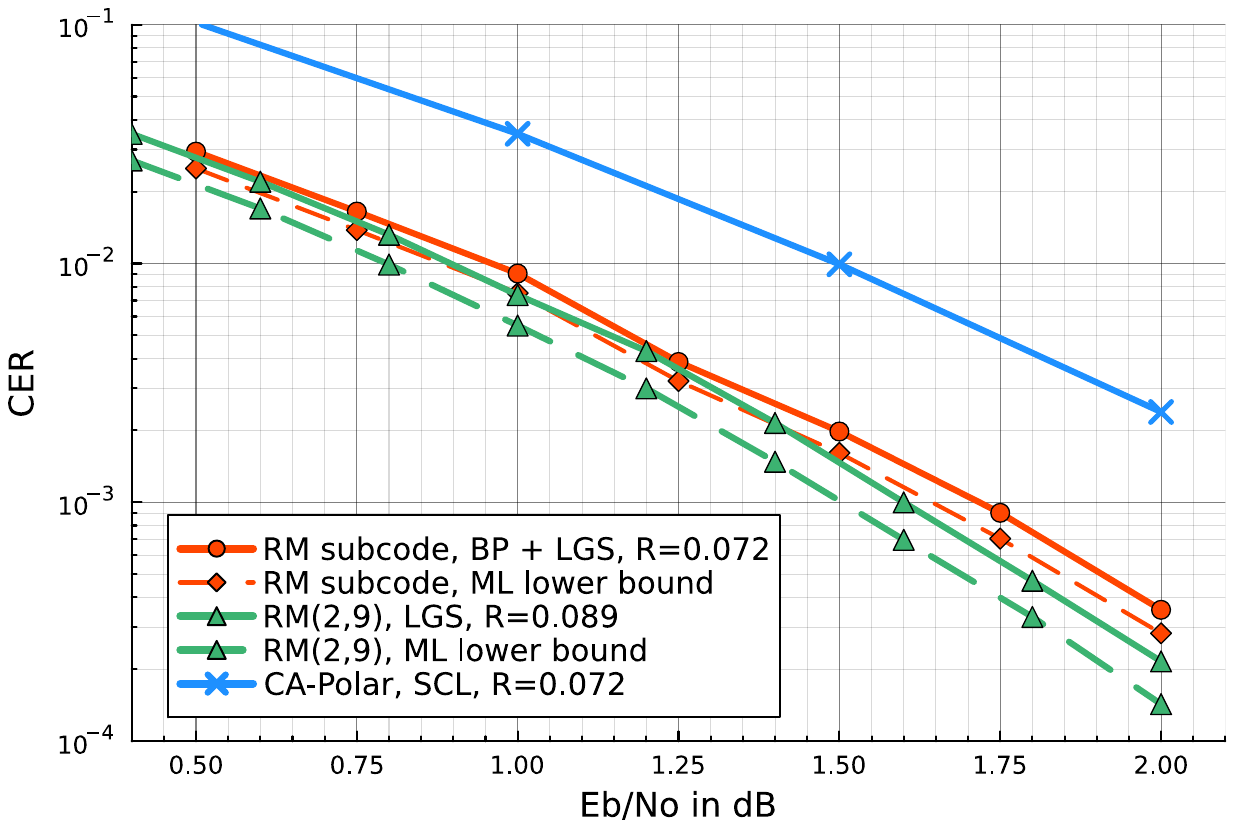}
    \caption{Comparison of codes with length $512$.}
    \label{fig:512}
\end{figure}
\begin{figure}[!t]
    \centering
    \includegraphics[width=0.75\columnwidth]{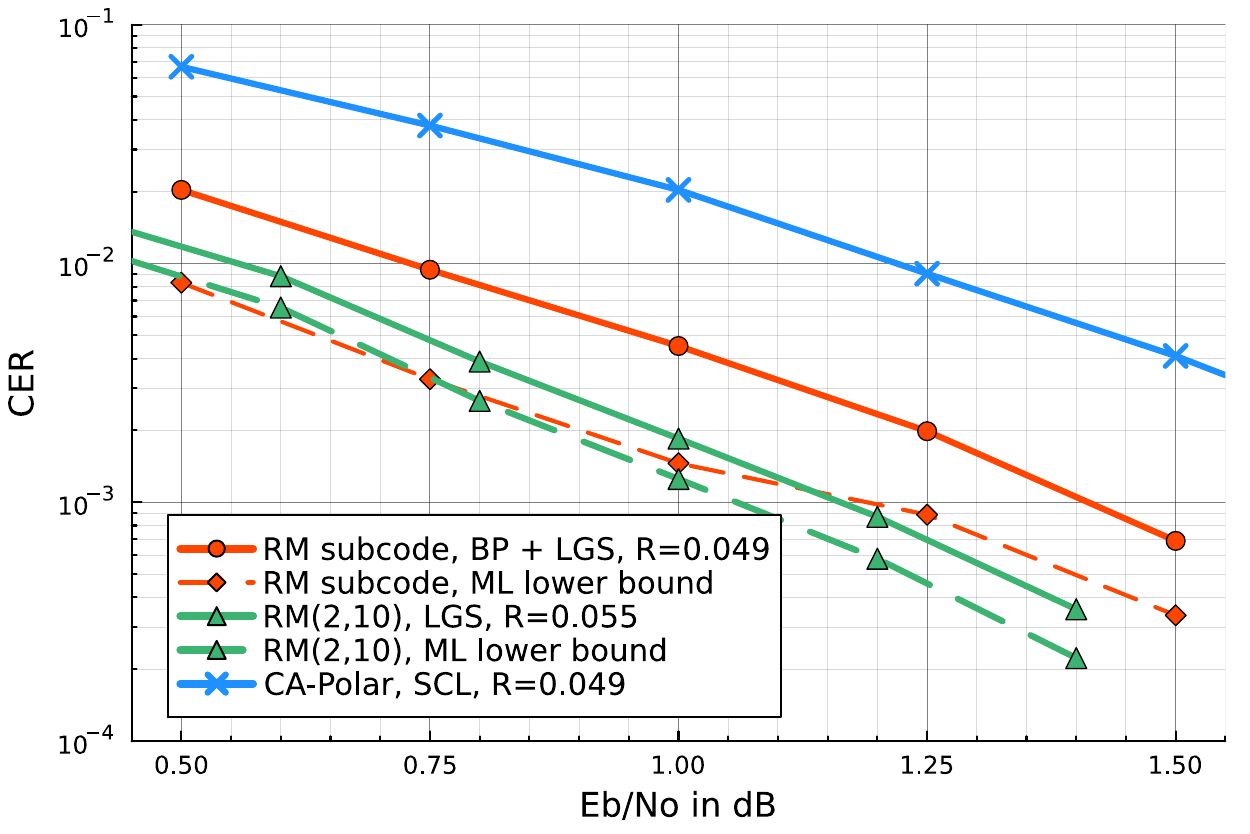}
    \caption{Comparison of codes with length $1024$.}
    \label{fig:1024}
\end{figure}

\section{Simulation Results} \label{sec:simulation_results}

We compare the CER (in the binary-input AWGN channel) of the recursive subproduct code $\code{2}{m}=\RM(1,m')^{\otimes [2,m]}$ with the second order RM code $\RM(2,mm')$ and the 5G NR CRC-aided Polar (CA-Polar) codes for $(m,m')=(4,2)$, $(3,3)$ and $(5,2)$. The block lengths corresponding to these choices of $(m,m')$ are $256$, $512$ and $1024$, respectively. In  all cases, the rate of the CA-Polar code is chosen to be equal to the rate of $\code{2}{m}$.

In all three simulations we observe that $\code{2}{m}$ performs better than the CA-Polar code and within $0.25$~dB of the RM code at \mbox{$\text{CER}\,=10^{-3}$}.
Note that the complexity of decoding $\code{2}{m}$, as presented in this work, is higher than RM and CA-Polar codes. 
However, our main objective is to highlight the fact that $\code{2}{m}$ can perform close to RM codes under near-ML decoding. 
We are yet to optimize the BP implementation (such as, intelligent scheduling to reduce the number of iterations, or use fewer generalized check nodes in order to reduce complexity). 

For decoding $\code{2}{m}$, we first use a BP decoder with iterations run till convergence to a valid codeword or till $100$ iterations, whichever is earlier.
A weighted variable node update rule is used (as in~\cite{LHP_ISIT_20,SNK_ISIT24}), where the messages intended to the variable nodes arising from the generalized check nodes are scaled. 
The messages arriving at the variable nodes from the nodes performing SISO decoding of $\RM(1,d) \otimes \RM(0,mm'-1-d)$ are multiplied by $0.006$, and the messages from nodes where $\RM(1,m')$ is decoded are multiplied by $0.2$. 
In case the BP output does not converge to a codeword, we perform hard-decision decoding on an information set of the BP output to estimate the codeword. 
This estimate is then used as the initial point $\p{c}^{(0)}$ in LGS decoding.
Also, we present a lower bound on the CER of $\code{2}{m}$ under maximum-likelihood (ML) decoding estimated using the same technique as in~\cite{DuS_IT_06,YeA_IT_20}.

For RM codes, we relied on the data available in~\cite{YeA_IT_20} and~\cite{Kam_TCOM_22} for CER under RPA-list decoding and LGS decoding, respectively.
We used the Matlab library~\cite{Matlab_Polar} for simulating 5G NR Polar codes (codes for uplink with $11$-bit CRC) under successive cancellation list (SCL) decoding with a list size of $64$. 

Fig.~\ref{fig:256} shows the performance of codes of length $256$, where $\code{2}{m}$ and the CA-Polar code have the same rate \mbox{$R=33/256$} and the RM code has rate \mbox{$R=37/256$}. The RPA-list decoder of the RM code~\cite{YeA_IT_20} has near-ML performance. The LGS decoder of $\code{2}{m}$ uses $P_{\sf LGS}=512$.
While $\code{2}{m}$ and the RM code have similar CER versus $E_b/N_o$, the difference in their code rates implies that the performance of $\code{2}{m}$ is better by approximately $10 \log_{10}\left( 37/33\right) \approx 0.49$~dB in terms of signal-to-noise ratio (${\rm SNR}=R E_b/N_o$).
Note that this particular RM subcode is optimal in the following sense (see Remark~\ref{rem:optimality_of_RM_subcodes}): among all subcodes of $\RM(2,8)$ of dimension $33$ that can be obtained by removing any $4$ rows of Hamming weight $64$ from the generator matrix of $\RM(2,8)$ this subcode has the least number of minimum weight codewords.

Fig.~\ref{fig:512} and~\ref{fig:1024} show the performance of codes with lengths $512$ and $1024$, respectively. In these figures, the CER of RM code under LGS decoding as well as an ML lower bound for RM codes is shown using the data available in~\cite{Kam_TCOM_22}. The LGS decoder of $\code{2}{m}$ uses $P_{\sf LGS}=2^{11}$ and $2^{13}$ for lengths $512$ and $1024$, respectively.
For length $512$, $\code{2}{m}$ has dimension $37$ (it is $46$ for the RM code), while for length $1024$ the dimension is $51$ (and $56$ for the RM code).

\section{Discussion} \label{sec:discussion}

We studied subcodes of $\RM(2,mm')$ obtained by removing the rows $x_ix_j$ from the generator matrix of $\RM(2,mm')$ where $i$ and $j$ belong to the same block. 
We are hopeful that these codes are optimal (in terms of number of minimum weight codewords) among subcodes of $\RM(2,mm')$ obtained by removing rows of weight $2^{mm'-2}$ from the latter's generator matrix; however, we do not have a proof yet. 
Also, we do not have a recurrence relation (similar to Lemmas~\ref{lem:recurrence_mdash_2} and~\ref{lem:recurrence_mdash_3}) to determine the weight distribution of codes with $m' \geq 4$.
It would be interesting to analyze the improvements (if any) in the CER performance of these codes when combined with precoding or CRC (as with Polar codes~\cite{Arikan_PAC_2019}).
There might be other systematic ways to construct good RM subcodes (with dimensions different from that of $\code{2}{m}$), such as removing rows $x_ix_j$ where $i$ and $j$ \emph{do not} belong to the same block. We do not know if such codes have a good CER performance too.



\appendices


\section{Proof of Lemma~\ref{lem:recurrence_mdash_2}} \label{app:lem:recurrence_mdash_2}

Note that any matrix in $\mathcal{B}_{2,m}$ can be obtained by appending two rows and two columns to a matrix from $\mathcal{B}_{2,m-1}$. 
%
Consider a $2m - 2 \times 2m - 2$ matrix $A \in \mathcal{B}_{2, m - 1}$. 
We first pad a row-column pair $\p{y}_{1}$ and $\p{y}_{1}^{T}$ followed by $[\p{y}_{2}~0] = \p{v}$ and $\p{v}^{T}$. The new diagonal entries are set to 0 after each round of padding, resulting in a $2m \times 2m$ matrix. 
After padding the first row-column pair $\p{y}_{1}$ and $\p{y}_{1}^{T}$, we have
\begin{equation*}
A_1 = 
\begin{bmatrix}
A & \p{y}_1^T \\
\p{y}_1 & 0  
\end{bmatrix}.
\end{equation*}
After padding $\p{v}$ and $\p{v}^T$, we arrive at
\begin{equation*}
A_2 = 
\begin{bmatrix}
A_1 & \p{v}^T \\
\p{v} & 0  
\end{bmatrix}
=
\begin{bmatrix}
A & \p{y}_1^T  & \p{y}_2^T \\
\p{y}_1 & 0  & 0 \\
\p{y}_2 & 0  & 0  
\end{bmatrix} \in \mathcal{B}_{2,m}.
\end{equation*}
We use the following fact from~\cite[Chapter~15]{macwilliams1977theory}: for any symplectic matrix, appending a row-column pair either preserves the rank or increases the rank by $2$. The rank is preserved if the appended row is linearly dependent on the rows of the original matrix, and the rank increases by $2$ otherwise. 
That is, $\rank(A_1) = \rank(A)$ if $\p{y}_1$ lies in the row span of $A$, i.e., $\p{y}_1 \in \rowspaceof(A)$, else $\rank(A_1) = \rank(A) + 2$.
Similarly, $\rank(A_2) = \rank(A_1)$ if \mbox{$\p{v} \in \rowspaceof(A_1)$}, else $\rank(A_2) = \rank(A_1) + 2$.
We denote the former scenario as `LD' (the appended row is linearly dependent on the rows of the original matrix), and the latter as `LI' (linearly independent).

Since $A_2$ is obtained from $A$ in two steps, we have four possible cases corresponding to $\p{y}_{1}$ and $\p{v}$ being LD or LI. If both are LD $\rank(A_2)=\rank(A)$, if both are LI we have $\rank(A_2)=\rank(A)+4$, else we have $\rank(A_2)=\rank(A)+2$. 
We now count the number of ways of arriving at a matrix $A_2$ with rank $h$ by considering these four cases one by one.


\emph{Case 1.} The first case is when both $\p{y}_{1}$ and $\p{v}$ are LD. Therefore, \mbox{$\rank(A) = h$} and we are going to determine $t_{1}$ in the recurrence stated in Lemma~\ref{lem:recurrence_mdash_2}. 
The number of choices of $\p{y}_1$ is $2^h$ since $\p{y}_1 \in \rowspaceof(A)$ and $\rank(A)=h$.
Now, we will count the number of choices for $\p{v}$ noting that $\p{v} \in \rowspaceof(A_1)$ and the last entry of $\p{v}$ must be $0$.
If $\p{y}_{1} = \p{0}$, all the $2^{h}$ vectors in $\rowspaceof(A_1)$ are valid choices for $\p{v}$ as their last entries are $0$. 
However, if $\p{y}_{1} \neq \p{0}$, then exactly $2^{h - 1}$ vectors in $\rowspaceof(A_1)$ have their last entry $0$. Here, we relied on the fact that for any subspace of $\Fb_2^{\ell}$ and any $i \in [\ell]$, the $i^\tth$ coordinate is $0$ either for all the vectors in the subspace or for exactly half of them. Therefore, $t_{1} = 1 \times 2^{h} + (2^{h} - 1) \times 2^{h - 1}$.

\emph{Case 2.} The second case is when $\p{y}_{1}$ is LD and $\p{v}$ is LI. Therefore, \mbox{$\rank(A) = h - 2$}, and the count derived in this case will contribute to one part of $t_2$, which will be denoted as $t_{2,1}$.
Similar to Case~1, there exist $2^{h-2}$ choices for $\p{y}_{1}$. 
To count the number of LI possibilities for $\p{v}$, we subtract the cases where $\p{v}$ is LD from its total possibilities.
Since $\p{v} \in \Fb_2^{2m - 1}$ and its last entry is 0, it admits $2^{2m - 2}$ total possibilities. Using the same argument as in Case~1, there are $2^{h - 2}$ and $2^{h - 3}$ possibilities such that $\p{v}$ is LD when $\p{y}_{1} = \p{0}$ and $\p{y}_{1} \neq \p{0}$, respectively. Putting all the arguments together, $t_{2,1} = 1 \times (2^{2m - 2} - 2^{h - 2}) + (2^{h - 2} - 1) \times (2^{2m - 2} - 2^{h - 3})$.

\emph{Case 3.} The penultimate case is when $\p{y}_{1}$ is LI and $\p{v}$ is LD. Therefore, \mbox{$\rank(A) = h - 2$}, and we are finding the other partial contribution to $t_{2}$, denote it as $t_{2,2}$. There are $2^{h - 2}$ possibilities such that $\p{y}_{1}$ is LD out of the $2^{2m - 2}$ total possibilities since $\p{y}_{1} \in \Fb_2^{2m - 2}$. Hence, there are $2^{2m - 2} - 2^{h - 2}$ options for $\p{y}_1$ that are LI. 
The rank of the matrix $A_1$ obtained after padding the row-column pair $\p{y}_{1}$ and $\p{y}_{1}^{T}$ is $h$. Since $\p{y}_{1} \neq \p{0}$, there are $2^{h - 1}$ possibilities for $\p{v}$ that are LD. Therefore, $t_{2,2} = (2^{2m - 2} - 2^{h - 2}) \times 2^{h - 1}$ and $t_{2} = t_{2,1} + t_{2,2}$.

\emph{Case 4.} The final case is when both $\p{y}_{1}$ and $\p{v}$ are LI. Therefore, $\rank(A) = h - 4$, and we are finding $t_{3}$. Given the reasoning presented in the previous case, there are $2^{2m - 2} - 2^{h - 4}$ ways in which $\p{y}_{1}$ can be chosen to be LI. The rank of the matrix obtained after padding the row-column pair \mbox{$\p{y}_{1} \neq \p{0}$} and $\p{y}_{1}^{T}$ is $h - 2$. Using the counting argument presented under the second case, there are $2^{2m - 2} - 2^{h - 3}$ ways such that $\p{v}$ can be chosen to be LI. Therefore, $t_{3} = (2^{2m - 2} - 2^{h - 4}) \times (2^{2m - 2} - 2^{h - 3})$.

This concludes the proof of Lemma~\ref{lem:recurrence_mdash_2}.

\section{Illustration of Proof of Lemma~\ref{lem:recurrence_mdash_3}} \label{app:lem:recurrence_mdash_3}

Consider a $3m - 3 \times 3m - 3$ matrix $A \in \mathcal{B}_{3, m - 1}$. 
We pad row-column pairs $\p{y}_{1}$ and $\p{y}_{1}^{T}$, $[\p{y}_{2}~0] = \p{v}$ and $\p{v}^{T}$, and $[\p{y}_{3}~0~0] = \p{w}$ and $\p{w}^{T}$ sequentially. The new diagonal entries are set to $0$ after each round of padding, resulting in a $3m \times 3m$ matrix $A_3 \in \mathcal{B}_{3,m}$. 
After padding the first row-column pair $\p{y}_{1}$ and $\p{y}_{1}^{T}$, we have
\begin{equation*}
A_1 = 
\begin{bmatrix}
A & \p{y}_1^T \\
\p{y}_1 & 0  
\end{bmatrix}.
\end{equation*}
After padding $\p{v}$ and $\p{v}^T$, we arrive at
\begin{equation*}
A_2 = 
\begin{bmatrix}
A_1 & \p{v}^T \\
\p{v} & 0  
\end{bmatrix}
=
\begin{bmatrix}
A & \p{y}_1^T  & \p{y}_2^T \\
\p{y}_1 & 0  & 0 \\
\p{y}_2 & 0  & 0  
\end{bmatrix}.
\end{equation*}
Finally, after padding $\p{w}$ and $\p{w}^T$, we have
\begin{equation*}
A_3 = 
\begin{bmatrix}
A_2 & \p{w}^T \\
\p{w} & 0  
\end{bmatrix}
=
\begin{bmatrix}
A & \p{y}_1^T  & \p{y}_2^T & \p{y}_3^T\\
\p{y}_1 & 0  & 0 & 0\\
\p{y}_2 & 0  & 0 & 0\\
\p{y}_3 & 0 & 0 & 0
\end{bmatrix}.
\end{equation*}
To obtain the recurrence as stated in Lemma~\ref{lem:recurrence_mdash_3}, we require $\rank(A_3)=h$.

As in the proof of Lemma~\ref{lem:recurrence_mdash_3}, $\p{y}_{1}$, $\p{v}$, and $\p{w}$ can be LI or LD giving rise to a total of $8$ cases. We show the approach for only $1$ of these $8$ cases, as the other cases can be analyzed in a similar manner and use ideas from proof of Lemma~\ref{lem:recurrence_mdash_2}.
We now consider the case where $\p{y}_{1}$, $\p{v}$, and $\p{w}$ are LD. Therefore, $\rank(A) = \rank(A_3) = h$, and we are finding the term $t_{1}$ in the recurrence. 

The number of choices for $\p{y}_{1} \in \rowspaceof(A)$ is $2^{h}$.
If $\p{y}_{1} = \p{0}$, all the $2^{h}$ vectors in $\rowspaceof(A_1)$ are valid choices for $\p{v}$ since the last entry of all these vectors is $0$. 
However, if \mbox{$\p{y}_{1} \neq \p{0}$}, then there are exactly $2^{h - 1}$ possibilities for $\p{v}$. 

We are now left with finding the number of possibilities for $\p{w}$. 
Towards this, we define two subspaces $S_{1} = \rowspaceof(A_2)$ and $S_{2} = \{\p{x} \in \Fb_2^{3m-1} : x_{3m-2}=x_{3m-1}=0 \}$. 
The number of possibilities for $\p{w}$ (or $\p{y}_3$) is exactly $|S_{1} \cap S_{2}| = 2^{\dim(S_{1} \cap S_{2})}$.
We know that 
\begin{equation*}
\dim(S_{1} \cap S_{2}) =  3m - 1 - \dim((S_{1} \cap S_{2})^{\perp}).  
\end{equation*}
Since 
\begin{align*}
\dim((S_{1} \cap S_{2})^{\perp}) &= \dim(S_{1}^{\perp} + S_{2}^{\perp}), \\
    &= \dim(S_{1}^{\perp}) + \dim(S_{2}^{\perp}) - \dim(S_{1}^{\perp} \cap S_{2}^{\perp})
\end{align*}
we are interested in the dimensions of $S_1^\perp$, $S_2^\perp$ and $S_1^\perp \cap S_2^\perp$.
Note that $\dim(S_{1}^{\perp}) = 3m - 1 - h$, $\dim(S_{2}^{\perp}) = 2$ and $\{\p{e}_{3m - 2},\p{e}_{3m - 1}\}$ is a basis for $S_2^\perp$. Thus,
\begin{equation*}
\dim(S_{1} \cap S_{2}) = h - 2 + \dim(S_{1}^{\perp} \cap S_{2}^{\perp}).
\end{equation*}
%
Now, we observe that 
\begin{align*}
\dim(S_{1}^{\perp} \cap S_{2}^{\perp}) = 
\begin{cases}
    2 & \text{ if } \p{y}_{1} = \p{y}_{2} = \p{0}, \\
    1 & \text{ if exactly one of } \p{y}_{1}, \p{y}_{2},\p{y}_{1} + \p{y}_{2} \text{ is } \p{0}, \\
    0 & \text{ if } \p{y}_{1}, \p{y}_{2} \text{ and } \p{y}_{1} + \p{y}_{2} \neq \p{0}.
\end{cases}
\end{align*}
We proceed with our counting by considering the following five subcases.

\emph{Subcase~1.} In this subcase we consider the situation $\p{y}_{1} = \p{y}_{2} = \p{0}$. The number of possibilities for $\p{y}_3$ is $2^h$. 
The total possibilities for the triple $(\p{y}_1,\p{y}_2,\p{y}_3)$ in this subcase is $2^h$.

\emph{Subcase~2.} Now consider $\p{y}_{1} = \p{0}$ and $\p{y}_{2} \neq \p{0}$. 
Clearly, $\p{y}_2$ can be chosen in $2^h - 1$ ways, and $\p{y}_3$ in $2^{h-1}$ ways. 
Thus, the number of possibilities for $(\p{y}_1,\p{y}_2,\p{y}_3)$ in this subcase is $(2^h-1)2^{h-1}$.

\emph{Subcase~3.} Consider $\p{y}_{1} \neq \p{0}$ and $\p{y}_{2} = \p{0}$. Here, $\p{y}_1$ has $2^h-1$ possibilities, $\p{y}_2$ has one possibility only, and $\p{y}_3$ has $2^{h-1}$ possibilities. Thus, $(\p{y}_1,\p{y}_2,\p{y}_3)$ has $(2^h-1)2^{h-1}$ possibilities. 

\emph{Subcase~4.} Consider $\p{y}_{1} = \p{y}_{2} \neq \p{0}$. 
Here, $\p{y}_1$ has $2^h-1$ possibilities, $\p{y}_2$ is equal to $\p{y}_1$, and since $\p{y}_1 + \p{y}_2 = \p{0}$, $\p{y}_3$ has $2^{h-1}$ possibilities, leading to $(2^h-1)2^{h-1}$ possibilities for $(\p{y}_1,\p{y}_2,\p{y}_3)$. 

\emph{Subcase~5.} Let both $\p{y}_1, \p{y}_2$ be non-zero and $\p{y}_{1} \neq \p{y}_{2}$.
Note that $\p{y}_1$ can be selected in $2^{h} - 1$ ways, $\p{y}_{2}$ in $2^{h - 1} - 2$ ways, and $\p{y}_3$ has $2^{h-2}$ possibilities. This subcase consists of $(2^{h} - 1)(2^{h - 1} - 2)2^{h - 2}$ possible values of $(\p{y}_1,\p{y}_2,\p{y}_3)$.

Adding the counts from all the five subcases together, we have $t_{1} = (1 \times 1 \times 2^{h}) + (3 \times (2^{h} - 1) \times 2^{h - 1}) + ((2^{h} - 1) \times (2^{h - 1} - 2) \times 2^{h - 2})$.
This completes the derivation of the term $t_1$ in the statement of Lemma~\ref{lem:recurrence_mdash_3}.

\section{Proof of Lemma~\ref{lem:dimension_of_Ua}} \label{app:lem:dimension_of_Ua}

\emph{Case 1.} Let us first consider the case when $\supp(\p{a})$ is a subset of a block. For concreteness, assume $\supp(\p{a}) \subset \Bs_1$. Observe that the first $m'$ entries of every vector in $U_{\p{a}}$ is $0$, and hence, $\dim(U_{\p{a}}) \leq mm'-m'$. 
To show that $\dim(U_{\p{a}}) \geq mm'-m'$, it is enough to show that the standard basis vectors $\p{e}_i$, \mbox{$i \in [mm'] \setminus [m']$} belong to $U_{\p{a}}$. 
To see that this is true, consider a matrix $B \in \mathcal{B}$ such that the $j^\tth$ row of $B$ is $\p{e}_i$ (and the $j^\tth$ column is $\p{e}_i^T$), and all other entries of $B$ are zero, where $j$ is any element in $\supp(\p{a})$. Clearly, $\p{a}B=\p{e}_i$.

\emph{Case 2.} We now consider the case when $\supp(\p{a})$ has non-empty intersection with two or more blocks.
To prove that $\dim(U_{\p{a}}) \leq mm' - 1$, we show that the vector $\p{a}$ used to generate $U_{\p{a}}$ belongs to the orthogonal subspace $U_{\p{a}}^{\perp}$. Therefore, $\dim(U_{\p{a}}^{\perp}) \geq 1$ which limits the dimension of $U_{\p{a}}$ to at most $mm' - 1$.
We first write $\p{a} = \left(\p{a}_{1}, \p{a}_{2}, \dots, \p{a}_{m} \right)$, where each $\p{a}_i$ is of length $m'$. For any $\p{u} \in U_{\p{a}}$, there exists a $B \in \mathcal{B}$ such that $\p{u} = \p{a}B$. Observe that $\p{u}\p{a}^T = \p{a}B\p{a}^T = \sum_{i=1}^{m} \sum_{j=1}^{m} \p{a}_{i}B_{i, j}\p{a}_{j}^{T}$. Using the facts $B_{i,i}=0$ and $B_{i,j}^T=B_{j,i}$, we have $\p{u}\p{a}^T = \sum_{i < j} \left( \p{a}_{i}B_{i, j}\p{a}_{j}^{T} + \p{a}_{j}B_{i,j}^T\p{a}_{i}^{T} \right) = 0$.

We are left with showing $\dim(U_{\p{a}}) \geq mm' - 1$. To simplify notation we now assume that $\supp(\p{a}) \cap \Bs_1, \supp(\p{a}) \cap \Bs_2 \neq \emptyset$, and in particular $\{1,m'+1\} \subset \supp(\p{a})$. The proof is similar for other cases. 
Note that $\mathcal{B}$ is a vector space, and the map $\varphi: \mathcal{B} \to U_{\p{a}}$ that sends $B$ to $\p{a}B$ is linear.
To complete the proof we will now identify a subspace $\mathcal{B}_{sub}$ of $\mathcal{B}$, of dimension $mm'-1$, such that the restriction of the linear map $\varphi$ to $\mathcal{B}_{sub}$ is one-to-one. 

We define  $\mathcal{B}_{sub}$ to be the collection of matrices $B \in \mathcal{B}$ such that all the entries except the following $2(mm'-1)$ entries are equal to zero: 
the last $mm' - m'$ entries of the $1^{\text{st}}$ row,  the first $m'$ entries of the $(m' + 1)^\tth$ row, and (since $B$ is symmetric), the last $mm' - m'$ entries of the $1^{\text{st}}$ column, and the first $m'$ entries of the $(m' + 1)^\tth$ column. 
Note that $\mathcal{B}_{sub}$ is a subspace of $\mathcal{B}$, and its dimension is $mm'-1$ since there is a free choice on the last $mm' - m'$ entries of the $1^{\text{st}}$ row and the entries $2,\dots,m'$ of the $(m' + 1)^\tth$ row, and the remaining entries of $B \in \mathcal{B}_{sub}$ are determined by these $mm'-1$ entries through the condition $B^T=B$. 
Using the facts that $1,m'+1 \in \supp(\p{a})$, it is straightforward to see that if $B \in \mathcal{B}_{sub}$ and $\varphi(B)=\p{a}B=\p{0}$ then necessarily $B=0$. This shows that the restriction of $\varphi$ to ${\mathcal{B}_{sub}}$ is one-to-one.

\section*{Acknowledgment}
The authors thank Aditya Siddheshwar for support towards the simulations.

\bibliographystyle{IEEEtran}
\bibliography{IEEEabrv,ISIT25}

\end{document}